\documentclass[a4paper,aps,twocolumn,prb,
showpacs,amsmath,amssymb,superscriptaddress,draft=false,nobibnotes]{revtex4-2}

\usepackage{graphicx}
\usepackage{dcolumn}
\usepackage{bm}
\usepackage{longtable}
\usepackage{color}
\usepackage[]{subfigure}
\usepackage[hidelinks]{hyperref}
\usepackage{color}

\begin{document}

\title{Hard X-ray photoemission 
spectroscopy of LaVO$_3$/SrTiO$_3$: Band alignment and electronic reconstruction }

\author{M. St\"ubinger}
\affiliation{Physikalisches Institut and W\"urzburg-Dresden Cluster of Excellence ct.qmat, Universit\"at W\"urzburg, Am Hubland, 97074 W\"urzburg, Germany}
\author{J. Gabel}
\affiliation{Physikalisches Institut and W\"urzburg-Dresden Cluster of Excellence ct.qmat, Universit\"at W\"urzburg, Am Hubland, 97074 W\"urzburg, Germany}
\affiliation{Diamond Light Source Ltd., Didcot, Oxfordshire OX11 0DE,
United Kingdom}
\author{P. Scheiderer}
\author{M. Zapf}
\author{M. Schmitt}
\author{P. Sch\"utz}
\author{B. Leikert}
\author{J. K\"uspert}
\author{M. Kamp}
\affiliation{Physikalisches Institut and W\"urzburg-Dresden Cluster of Excellence ct.qmat, Universit\"at W\"urzburg, Am Hubland, 97074 W\"urzburg, Germany}
\author{P.K. Thakur}
\author{T.-L. Lee}
\affiliation{Diamond Light Source Ltd., Didcot, Oxfordshire OX11 0DE,
United Kingdom}

\author{P. Potapov}
\author{A. Lubk}
\author{B. B\"uchner}
\affiliation{Leibniz Institute for Solid State and Materials Research and W\"urzburg-Dresden Cluster of Excellence ct.qmat, 01069 Dresden, Germany}

\author{M. Sing}
\author{R. Claessen}
\affiliation{Physikalisches Institut and W\"urzburg-Dresden Cluster of Excellence ct.qmat, Universit\"at W\"urzburg, Am Hubland, 97074 W\"urzburg, Germany}
\date{\today}

\begin{abstract}

The heterostructure consisting of the Mott insulator LaVO$_3$ and the 
band insulator SrTiO$_3$ is considered a promising candidate for future 
photovoltaic applications. Not only does the (direct) excitation gap of LaVO$_3$ 
match well the solar spectrum, but its correlated nature and predicted built-in 
potential, owing to the non-polar/polar interface when integrated with SrTiO$_3$,
also offer remarkable advantages over conventional solar cells. 
However, experimental data beyond the observation of a thickness-dependent metal-insulator 
transition is scarce and a profound, microscopic understanding 
of the electronic properties is still lacking. By means of soft and hard X-ray 
photoemission spectroscopy as well as resistivity and Hall effect 
measurements we study the electrical properties, band bending, 
and band alignment of LaVO$_3$/SrTiO$_3$ heterostructures. We find a critical LaVO$_3$ thickness of five unit cells, 
confinement of the conducting electrons to exclusively Ti 3$d$ states at the 
interface, and a potential gradient in the film. From these findings we conclude 
on electronic reconstruction as the driving mechanism for the formation of the metallic interface in 
LaVO$_3$/SrTiO$_3$.

\end{abstract}

\maketitle

\section{Introduction}
\label{Introduction}

\begin{figure}[!htb]

\includegraphics[width = 0.45\textwidth]{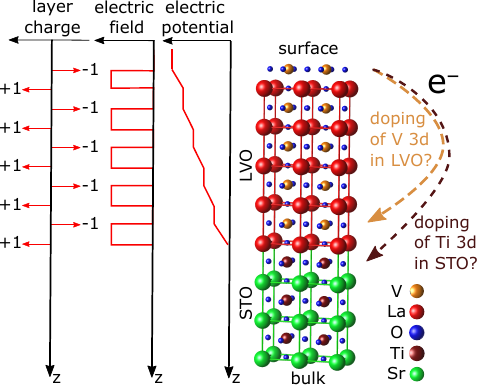}

\caption{
\label{RE}
In the electronic reconstruction model the polar discontinuity at the 
interface of the LVO/STO heterostructure leads to a potential built-up in LVO. 
It can be compensated by transferring electrons from the 
surface to the interface. In LVO/STO, the 3$d$ shells of both Ti as well as V 
could host the electrons.}
\end{figure}

Transition-metal oxides with strong electronic correlations are of great 
interest in condensed-matter physics due to their vast amount of unique 
properties. Among them are high-temperature superconductivity, colossal 
magnetoresistance, and metal-to-insulator transitions, functionalities, 
that offer a versatile platform to build on future 
devices \cite{morosan_strongly_2012,uehara_percolative_1999, 
takagi_emergent_2010,ahn_electric_2003,yajima_heteroepitaxial_2011, 
scheiderer_tailoring_2018}. Specifically, the strong Coulomb repulsion of the 
rather localized valence electrons in transition-metal oxides with incomplete 
valence shells can lead to a Mott insulating ground state \cite{fujimori_electronic_2001}. Mott insulators with a 
charge gap in the regime of visible light have been 
proposed to be promising candidates for photovoltaic applications due to their 
favorable absorption 
properties regarding the 
solar spectrum \cite{assmann_oxide_2013,zhang_high-quality_2017}. Furthermore, Mott insulators as photovoltaic materials have 
some intriguing advantages over conventional semiconductors. By so called 
impact ionization more than one electron per 
incoming photon can be excited before thermalization. This is due to 
electron-electron interactions being much faster than electron-phonon 
interactions, which are the dominant mechanism for thermalization in 
conventional semiconductors \cite{werner_role_2014, manousakis_photovoltaic_2010}. Besides, a fast separation 
of photogenerated electron-hole pairs may be achieved by the antiferromagnetic 
background of many Mott insulators, which acts as an energy buffer for 
photoexcited carriers on femtosecond time 
scales \cite{eckstein_ultrafast_2014}.

The system to be studied here, LaVO$_3$ (LVO), is such an 
antiferromagnetic Mott insulator with an optical gap of 
1.1\,eV \cite{wang_device_2015, arima_variation_1993}. Its usage for photovoltaic applications can be 
further optimized by integrating this material as thin film with the band 
insulator SrTiO$_3$ (STO) as substrate into a heterostructure, since the 
polar discontinuity at the interface may lead to a potential gradient in the 
LVO 
film, which would help to separate the photogenerated electron hole-pairs. 
While\textemdash in the ionic limit\textemdash STO consists of charge-neutral 
layers along the [001] 
direction, LVO is built up by alternating positive (LaO)$^+$ and negative 
(VO$_2$)$^-$ layers. The polar discontinuity between STO and LVO induces an 
electrostatic potential which diverges with LVO film thickness 
(see Fig.~\ref{RE}). 
Above a critical film thickness, the interface turns 
conductive \cite{hotta_polar_2007}. Density-functional theory (DFT) calculations \cite{assmann_oxide_2013} 
show that in 
such a situation the internal potential gradient in the LVO/STO heterostructure 
persists and can be exploited to separate photogenerated 
electron-hole pairs. This and the metallic interface and surface 
serving as electrodes to extract the charge carriers, renders LVO/STO 
an interesting candidate for oxide-based solar cells \cite{assmann_oxide_2013, wang_device_2015, zhang_high-quality_2017, jellite_investigation_2018, goyal_persistent_2020}. 

Up to now, an in-depth understanding of this heterostructure regarding the 
emerging metallicity and the role of the internal potential is still lacking. 
For future device applications in photovoltaics, however, it is of great 
importance to comprehend 
the complete electronic band diagram and the role of charges in this system to 
tailor the desired properties. In our study, we investigate the charge transfer 
and the electronic band alignment in LVO/STO by transport measurements and photoemission spectroscopy. We show that conductivity is not 
induced by oxygen vacancies in the substrate but due to electronic 
reconstruction. The STO Ti~3\textit{d} bands host the electrons giving rise to 
to the metallicity while there is no sign for metallic V~3\textit{d} states at 
the Fermi energy.

\section{Experimental Details}

To prepare STO substrates with TiO$_2$ termination, the 
substrates were rinsed in deionized water, etched with hydrofluoric acid and 
subsequently annealed in oxygen \cite{koster_quasi-ideal_1998}. The LVO thin 
films were grown by pulsed laser deposition at a substrate temperature of 
550\,$^{\circ}$C and an oxygen pressure of 10$^{-6}$\,mbar. The LVO 
was ablated from a polycrystalline LaVO$_4$ target. The pulsed KrF excimer laser 
was set to a repetition rate of 1\,Hz and a laser fluency of 1.5\,J/cm$^2$. 
Reflection high-energy electron diffraction (RHEED) was employed to monitor the 
film growth in real time.

Atomic resolution scanning transmission electron microscopy (STEM) of the interface
was performed in a probe-corrected FEI Titan$^3$ microscope equipped with a Gatan Tridiem energy filter and operating at 300kV acceleration voltage.

Photoemission spectroscopy experiments were performed at the beamline I09 at 
Diamond Light Source. Soft x-rays with the photon energy set to $h\nu = 460.05$ 
eV and 
$517.7$ eV were used for measurements at the Ti~$L$ and V~$L$ resonance, 
respectively, and a sample temperature of 60\,K. Additionally, hard X-ray 
photoemission spectroscopy (HAXPES) with  
a photon energy of 3\,keV was used at room temperature. The EW4000 photoelectron analyzer 
(VG Scienta, Sweden) was equipped with a wide-angle acceptance lens. To 
prevent overoxidation of the LVO films, a portable ultrahigh vacuum (UHV) 
suitcase with a base 
pressure in the 10$^{-9}$\,mbar range was used to ship the samples to Diamond.

Temperature-dependent transport measurements were performed with a Physical 
Property Measurement System (Quantum Design, USA). The samples were 
electrically contacted by ultrasonic bonding with Al wires arranged in van der 
Pauw geometry.

\section{Characterization of structural, chemical and electronic properties}
\noindent

\begin{figure*}[!htb]

\includegraphics[width = 0.98\textwidth]{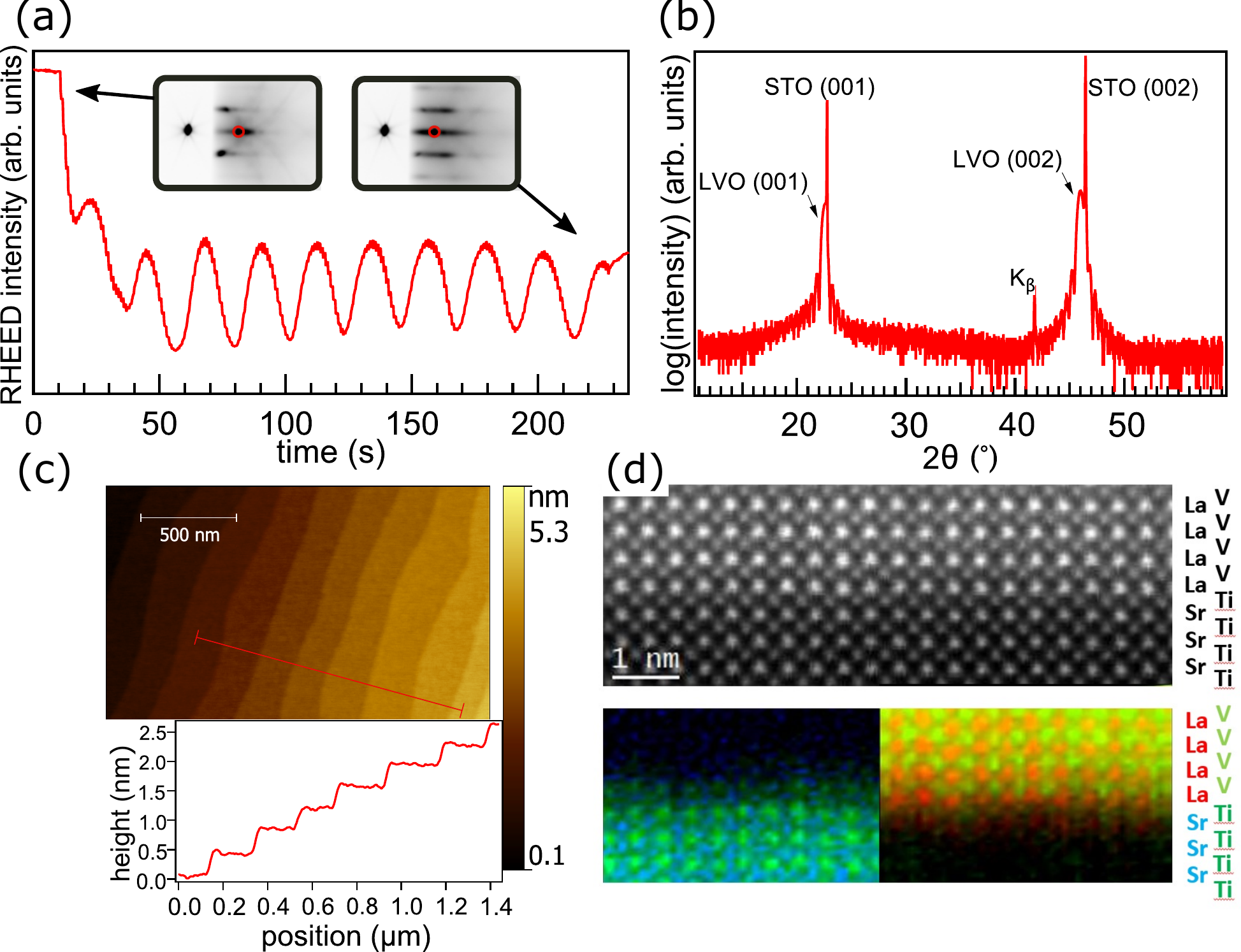}

\caption{
\label{growth}
(a) Monitoring of film growth by RHEED. Regular intensity oscillations of the 
specular reflex indicate layer-by-layer growth. The RHEED patterns confirm that 
the substrate and the film surface after the growth of LVO are atomically 
smooth. (b) X-ray diffractogram of a phase pure 50\,uc LVO film on STO. (c) AFM 
image of a 10\,uc LVO/STO sample along with a height profile signaling good 
quality of the sample surface. (d) Scanning transmission electron microscopy of a 30\,uc LVO/STO sample. (top) HAADF-STEM image showing the interface structure. (bottom) Chemical maps of the same region for substrate and film elements from STEM-EELS. The images confirm a chemically sharp  interface and a good structural film quality.}

\end{figure*}

Figure \ref{growth} shows the growth characterization of the LVO films. 
Figure \ref{growth}(a) displays the intensity modulations of the specular RHEED beam 
(see red circles in inset) during the film growth. The regular intensity 
oscillations signal layer-by-layer growth, while the RHEED pattern at the 
end of the growth confirms an atomically smooth film surface. In Fig. \ref{growth}(b) the X-ray 
diffraction (XRD) pattern 
of a sample with a 50 unit cells (uc) thick LVO film  
exhibits peaks corresponding only to a (pseudo)cubic (00l) LVO film as well as substrate peaks, 
demonstrating the phase purity of the sample. The atomic force microscopy (AFM) 
image in Fig. 
\ref{growth}(c) reveals the surface topograpy of a 10\,uc LVO/STO sample. One 
recognizes the terraced structure reflecting the miscut STO substrate 
whose surface is covered by a smooth film of uniform thickness. Also shown is a 
height profile along the red line indicated in the AFM image. 
Figure \ref{growth}(d, top) shows the interface of a 30 uc LVO/STO sample obtained in high-angle annular dark field (HAADF) STEM mode (contrast of atomic columns roughly proportional to squared atomic number) with the layer sequence matching exactly the schematic sequence in Fig. \ref{RE}. Figure \ref{growth}(d, bottom) displays chemical maps of the same region for substrate and film elements from STEM-EELS. The observable slight signal intermixing such as Ti traces at the adjacent La and V layers or La traces at the adjacent Ti and Sr layers is mainly caused by electron beam spreading within the sample due to scattering delocalization. Therefore, within the precision of the STEM technique, there were no hints for an off-stoichiometry or intermixing, and the LVO/STO interface was found structurally and chemically well defined on an atomic level.

\begin{figure}[!htb]

\includegraphics[width = 0.45\textwidth]{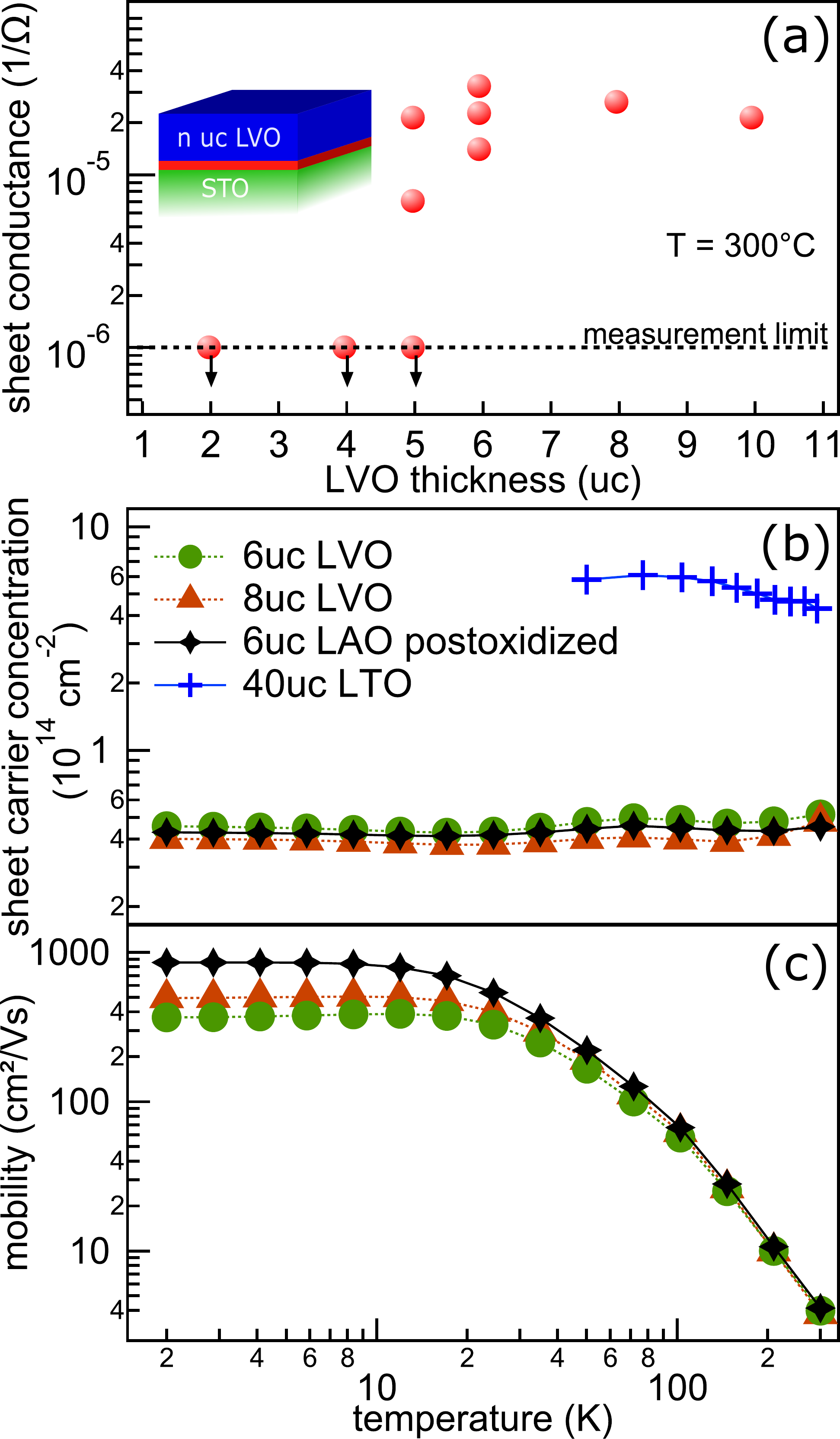}

\caption{
\label{SR}
(a) Sheet conductance of the LVO/STO heterostructure at room temperature for 
varying film thickness. Samples with LVO thicknesses above the critical value 
of 
five unit cells show metallic behavior. (b) Sheet carrier concentrations of metallic LVO/STO as well as LAO/STO 
samples and of the doped Mott insulator LTO. The behavior of the LVO 
films resembles that of post-oxidized LAO/STO. (c) Mobilities of metallic LVO/STO 
as well as LAO/STO samples.}
\end{figure}

After having confirmed the structural and chemical quality of the grown films, we now 
investigate the transport properties with resistivity and Hall effect 
measurements. Although both components are insulators, the LVO/STO 
heterostructure can become conductive as depicted in the model of intrinsic 
electronic reconstruction in Fig. \ref{RE}: According to this model, the built-in potential can 
be compensated by transferring half an electron per two-dimensional unit cell 
from the surface to the interface where transition metal 3\textit{d} orbitals 
can host the electrons, resulting in a metallic interface. This mechanism has 
been well investigated for the related LAO (LaAlO$_3$)/STO system 
\cite{nakagawa_why_2006, liu_origin_2013, thiel_tunable_2006}. 
Previous DFT calculations as 
well as hard X-ray core-level photoemission spectroscopy of 
trilayer LAO/LVO/LAO grown on STO suggest that charge transfer may 
involve not only Ti but also V sites and the involvement of the latter may lead to the emergence of 
strong electron correlations \cite{assmann_oxide_2013,takizawa_spectroscopic_2009}.
Hence, for device applications, it is crucial to elucidate where exactly the 
transferred electrons reside: the Ti~3\textit{d} states of 
the STO substrate, resulting in a slightly $n$-doped semiconducting interface 
layer, or the V~3\textit{d} states via self-doping within the LVO film, 
inducing a band-filling controlled Mott-insulator-to-metal transition 
and hence a correlated metal at the interface. However, one should 
keep in mind that the electronic reconstruction model in its ideal form is 
still highly disputed and that additional, extrinsic mechanisms for the 
generation of the two-dimensional electron system (2DES) like interfacial 
intermixing 
 \cite{willmott_structural_2007,qiao_epitaxial_2011}, oxygen vacancies 
 \cite{herranz_high_2007, zhong_polarity-induced_2010} or surface adsorbates 
 \cite{xie_control_2011,bristowe_surface_2011,scheiderer_surface-interface_2015} 
might be at work.

Figure~\ref{SR}~(a) shows the room temperature sheet conductance of LVO/STO samples 
with different film thicknesses. The insulator-to-metal transition occurs at a 
film thickness of 5\,uc. Films thinner (thicker) than 5\,uc are insulating 
(conductive).
A more intricate situation is found for the 
transition at 5\,uc, where we fabricated and measured four samples in total.
Two of the samples are metallic with a sheet conductance of the order of 
10$^{-5}$/Ohm, which is comparable with the thicker samples. The other two 
are insulating just like the films with a lower thickness. This behavior suggests
that at the critical thickness the sample's metallicity can be sensitive to
small variations in the substrate and film quality.

Our result of an insulator-to-metal transition at a film thickness of 5\,uc 
is consistent with the findings by Hotta \textit{et al.}, who reported a 
transition between 4 and 5\,uc \cite{hotta_polar_2007}. In a recent study, Hu 
\textit{et al.} pointed out that the mechanism behind the conductivity 
strongly depends on the growth parameters, especially the growth rate and the 
substrate temperature \cite{hu_elucidating_2019}. Authors who used a higher 
substrate temperature (up to 750$^{\circ}$C) reported 
oxygen-vacancy-induced conductivity \cite{rotella_two_2015}. This is mainly due 
to oxygen vacancies in STO that act as electron donors, rendering the 
substrate metallic \cite{kalabukhov_effect_2007}. Such metallicity solely based 
on STO is well established in literature 
\cite{santander-syro_two-dimensional_2011, meevasana_creation_2011, 
rodel_orientational_2014, mckeown_walker_control_2014, 
walker_carrier-density_2015, dudy_situ_2016, aiura_photoemission_2002, 
wang_anisotropic_2014}. Other groups used a 
lower substrate temperature (around 600$^{\circ}$C) and attributed the 
observed conductivity to the intrinsic mechanism of electronic 
reconstruction \cite{hotta_polar_2007,he_metal-insulator_2012}. Therefore, to 
suppress charges donated by oxygen vacancies in the 
substrate it is critical to cautiously adjust the growth parameters. Note 
that post-annealing in high oxygen pressures to get rid of oxygen vacancies is 
not possible as the thermodynamically more stable LaVO$_4$ phase would quickly 
form. We rather used a particularly low substrate 
temperature of 550$^{\circ}$C to minimize the creation of oxygen vacancies.

Figure~\ref{SR}~(b) depicts the temperature-dependent sheet carrier 
concentrations for different oxide heterostructures. The carrier 
concentrations of the metallic LVO/STO samples are essentially constant over 
the 
whole temperature range with a value of about $5 \cdot 10^{13}$\,cm$^{-2}$. 
This value is orders of magnitude lower than those for samples loaded with 
oxygen vacancies \cite{hu_elucidating_2019}. It is comparable to the 
values reported for intrinsic transport 
behavior \cite{hotta_polar_2007,he_metal-insulator_2012,hu_elucidating_2019} but 
much lower than the theoretical value of 3.28$\cdot$10$^{14}$\,cm$^{-2}$ that is 
expected when half an electron per unit cell is transferred to the interface in 
the ideal electronic reconstruction scenario. One also should note that 
variations of the cation stoichiometry likewise may play a decisive role in the 
transport properties. In a recent study of 50\,uc thick LVO films grown 
on STO, it has been found that a slight La-deficiency is mandatory 
for a metallic interface \cite{tomar_conducting_2020}.

Interestingly, the temperature-dependent carrier concentrations 
[Fig.~\ref{SR}~(b)] and mobilities [Fig.~\ref{SR}~(c)] of the metallic 
LVO/STO samples are almost identical to those of a post-annealed LAO/STO 
sample, whose transport properties have been shown to result from an intrinsic 
mechanism \cite{gabel_disentangling_2017}. These observations suggest 
that conductivity in the LVO/STO samples is not induced by oxygen vacancies in 
the substrate but is of intrinsic origin. Furthermore, the type of 
mobile carriers seems to be the same as in LAO/STO samples, i.e., electrons in 
Ti~3\textit{d} bands of STO (rather than electrons in V~3\textit{d} bands of 
LVO) are responsible for the conductivity. Indeed, electron doping of STO by an 
intrinsic electronic reconstruction type of mechanism creates a 2DES with a 
small density of charge carriers that are hosted by an almost empty 
Ti~3\textit{d} shell corresponding to a \textit{d}$^{0+\delta}$ configuration. 
On the contrary, electron doping of V~3\textit{d} states in LVO would result 
in  a slightly doped Mott insulator with a \textit{d}$^{2+\delta}$ occupancy. 
Such a Mott insulator, slightly $n$-doped into the metallic regime, would have 
a much higher charge carrier density since it is close to half band-filling. To 
compare our results for LVO/STO with those of a doped Mott insulator, Fig.~\ref{SR}~(b) also depicts the carrier concentration of a 40\,uc thick 
LaTiO$_3$ (LTO) film which is \textit{p}-doped by excess oxygen just into the 
metallic regime \cite{scheiderer_tailoring_2018}. For this doped 
Mott insulator we measure a much higher carrier concentration than for 
LVO/STO. Thus, with the charge carrier concentration of LVO/STO being almost 
identical to that of LAO/STO but way smaller than that for doped LTO we 
conclude that the metallicity of our LVO/STO samples is solely based on Ti 3$d$
states of the STO substrate.


\begin{figure*}[!htb]

\includegraphics[width = 0.99\textwidth]{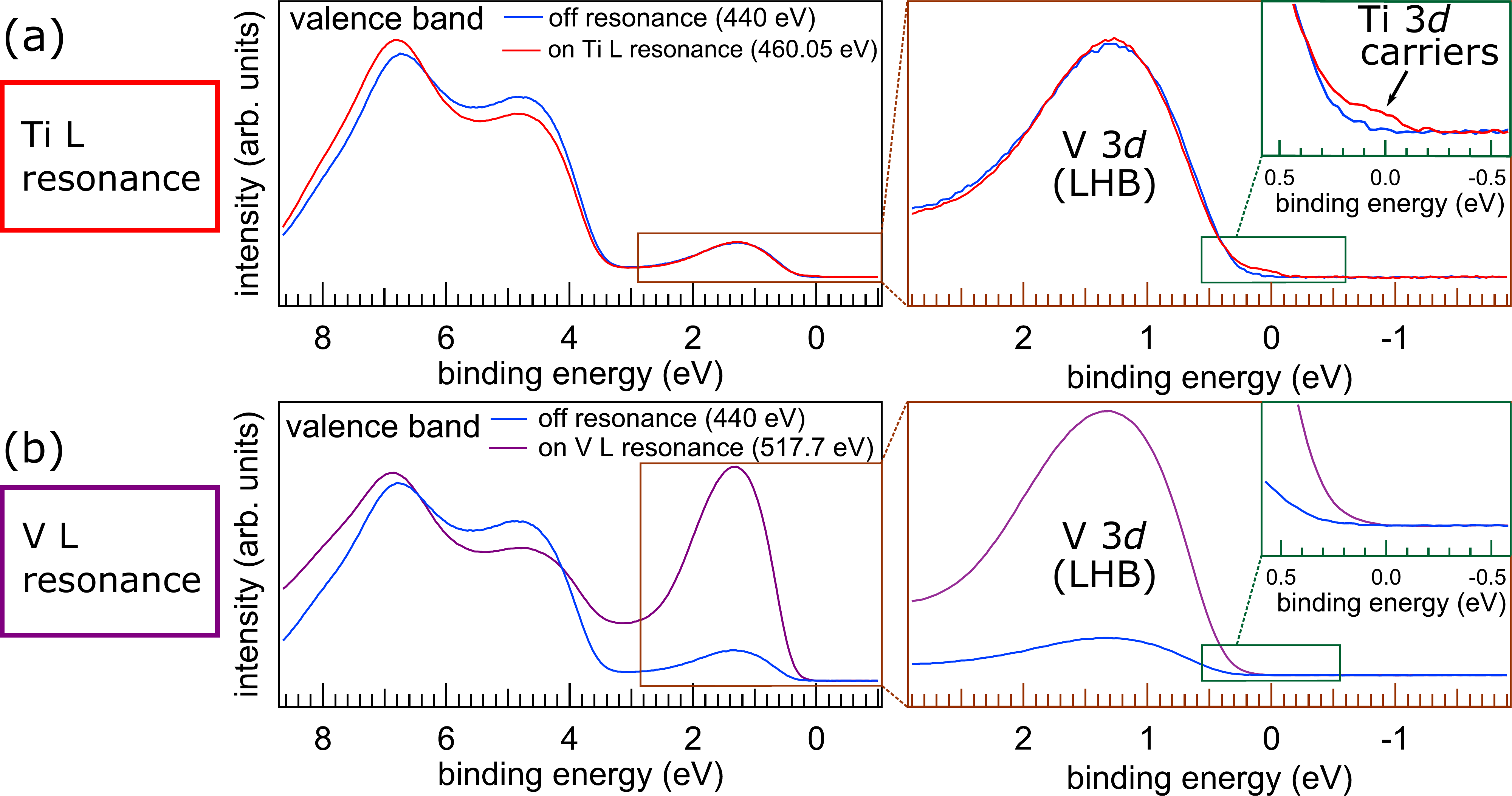}

\caption{
\label{RESPES}
(a) ResPES at the Ti~$L$ edge of a 
5\,uc LVO film on STO. On resonance, there is spectral weight at the 
Fermi energy, indicating metallic Ti~3\textit{d} charge carriers (marked by arrow). (b) ResPES at 
the V~$L$ edge of the same sample. On resonance, the lower Hubbard 
band (marked as V~3\textit{d} LHB) below the Fermi energy strongly resonates but there is no spectral weight 
at the Fermi energy to be seen, indicating the absence of metallic 
V~3\textit{d} charge carriers.}

\end{figure*}

To reveal the elemental and orbital character of the charge carriers, being 
responsible for the conductivity in this system, we employed resonant 
photoemission spectroscopy (ResPES) in the soft x-ray regime. In the case of 
LAO/STO, it is well established that the Ti~3$d$ charge 
carriers at $E_F$, forming the buried 2DES, can be studied by this technique 
despite the small carrier concentrations \cite{gabel_disentangling_2017, 
berner_direct_2013,drera_spectroscopic_2011}. Accordingly, we measured the 
valence band spectra of a 5\,uc LVO/STO sample with a conducting interface, both at the Ti~$L$ and V~$L$ edges. 
The sample was aligned parallel to the $\Gamma$--$X$ direction. The spectra were integrated over an angular range of $\pm23^\circ$ around normal emission (NE) parallel to the analyzer slit and $\pm0.15^\circ$ perpendicular to the analyzer slit.

Figure~\ref{RESPES}~(a) shows the valence band spectrum recorded
on the Ti~$L$ resonance together with an off-resonance spectrum. On resonance, 
there is spectral weight discernible at the Fermi energy. This is a clear 
indication of Ti~3\textit{d} charge carriers, corroborating our reasoning from 
the transport data.

Figure~\ref{RESPES}~(b) depicts the V resonance for the same sample. Here, the 
spectral weight related to the lower Hubbard band (LHB) at around 1.5\,eV below 
the Fermi energy strongly resonates when the photon energy is tuned to the 
V~$L$ resonance. In contrast to the Ti~$L$ resonance, however, no 
spectral weight is seen at the Fermi energy, indicating the absence of metallic
V~3\textit{d} charge carriers. Again, we arrive at the conclusion that the 
conductivity in LVO/STO is solely related to Ti~3$d$ charge carriers.

\section{Chemical and band structure information from HAXPES}

\begin{figure}[]

\includegraphics[width = 0.9\linewidth]{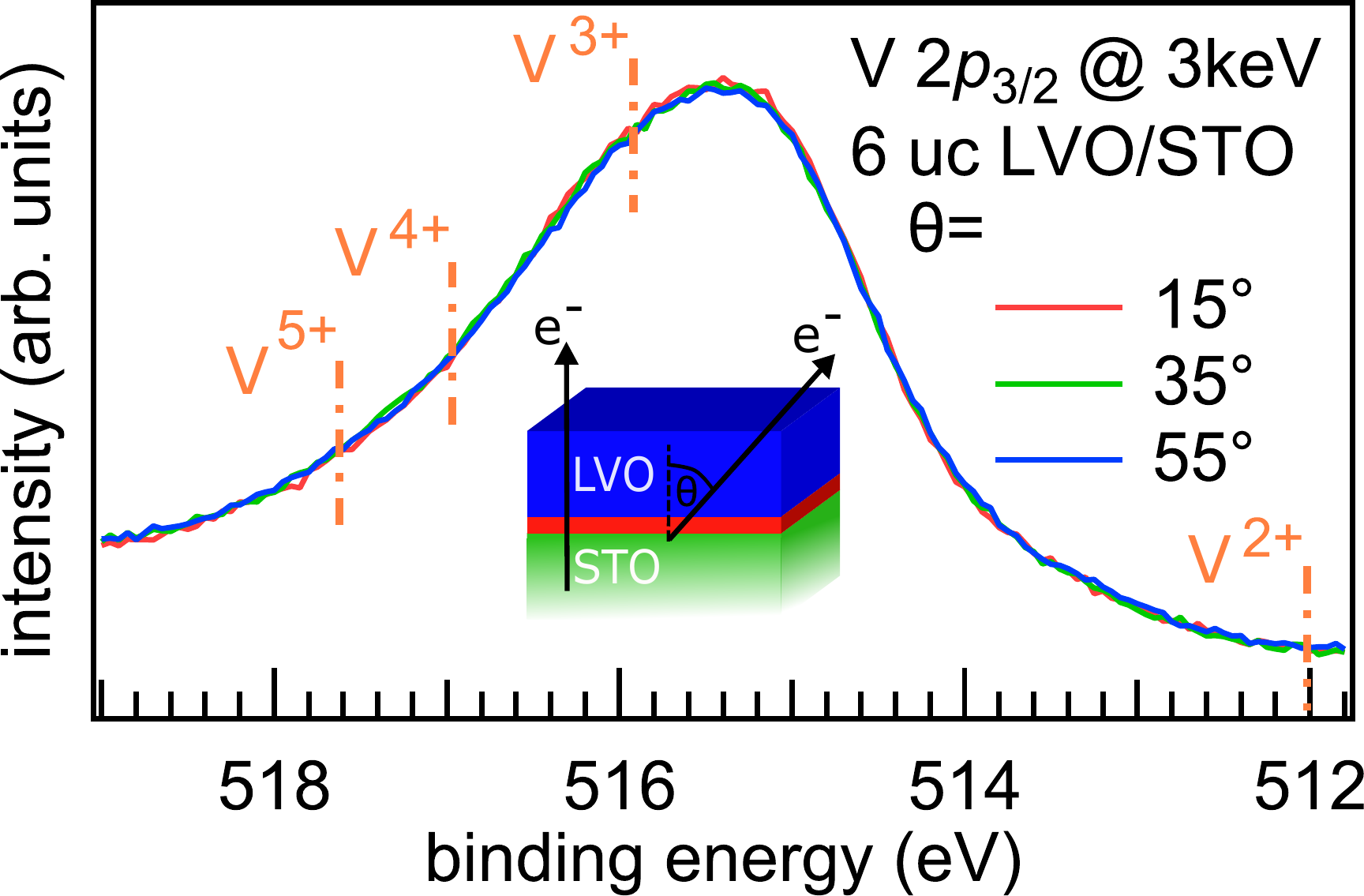}

\caption{
\label{core}
V~2\textit{p}$_{3/2}$ photoemission spectra of a 6\,uc LVO/STO sample for 
different electron emission angles. The spectra measured with higher angles off 
normal emission are more surface sensitive. The fact that there is no 
significant angle-dependence suggests a homogeneous distribution of vanadium 
with a uniform valence over the whole film and  
accordingly the absence of electron doping into V~3$d$ states.}
\end{figure}

To corroborate the results of our transport and Res\-PES data and to get the full 
band 
picture of the LVO/STO heterostructure, we 
investigated the electronic structure by HAXPES.  

\subsection{Homogeneous valence distribution and potential gradient in the 
LaVO$_3$ film from core-level analysis}

As Wadati \textit{et al.} pointed out, LVO is not stable in air since the 
surface will oxidize, leading to a V\,2\textit{p} photoemission spectrum 
with spectral weight shifted towards higher binding energies, reflecting a 
change from the nominal V$^{3+}$ valence of stoichiometric LVO to a higher V 
valence \cite{wadati_characterization_2007}. This is possible because the 
transition metal vanadium possesses oxidation states ranging from 0 to 
5+. Vanadium oxides in general---when exposed to air---tend to further oxidize at their 
surface, in this case to V$^{4+}$ and V$^{5+}$ \cite{hryha_stoichiometric_2012}. For these reasons, we used a UHV 
suitcase for shipping the samples from the PLD chamber to the synchrotron to preserve 
the stoichiometric LVO with its intrinsic electronic structure.

Figure \ref{core} depicts the V\,2\textit{p}$_{3/2}$ core level spectra of a 
6\,uc LVO/STO sample at 3\,keV photon energy collected at different electron emission 
angles $\theta$. In comparison to reported binding energies for different V 
valences, the spectrum appears most closely attributable to the 
V$^{3+}$ oxidation 
state \cite{hryha_stoichiometric_2012}. The broadening is mainly due to the 
partially filled \textit{d}-shell in LVO. The two 
V~3\textit{d} electrons couple to the core hole, leading to a multiplet 
splitting of the photoemission line \cite{zimmermann_strong_1998}.

The spectra for different emission angles, i.e., recorded with different 
probing depths, are essentially identical, meaning that the distribution 
of V valence states is homogeneous throughout the film. 

\begin{figure}[!htb]
\includegraphics[width = 0.85\linewidth]{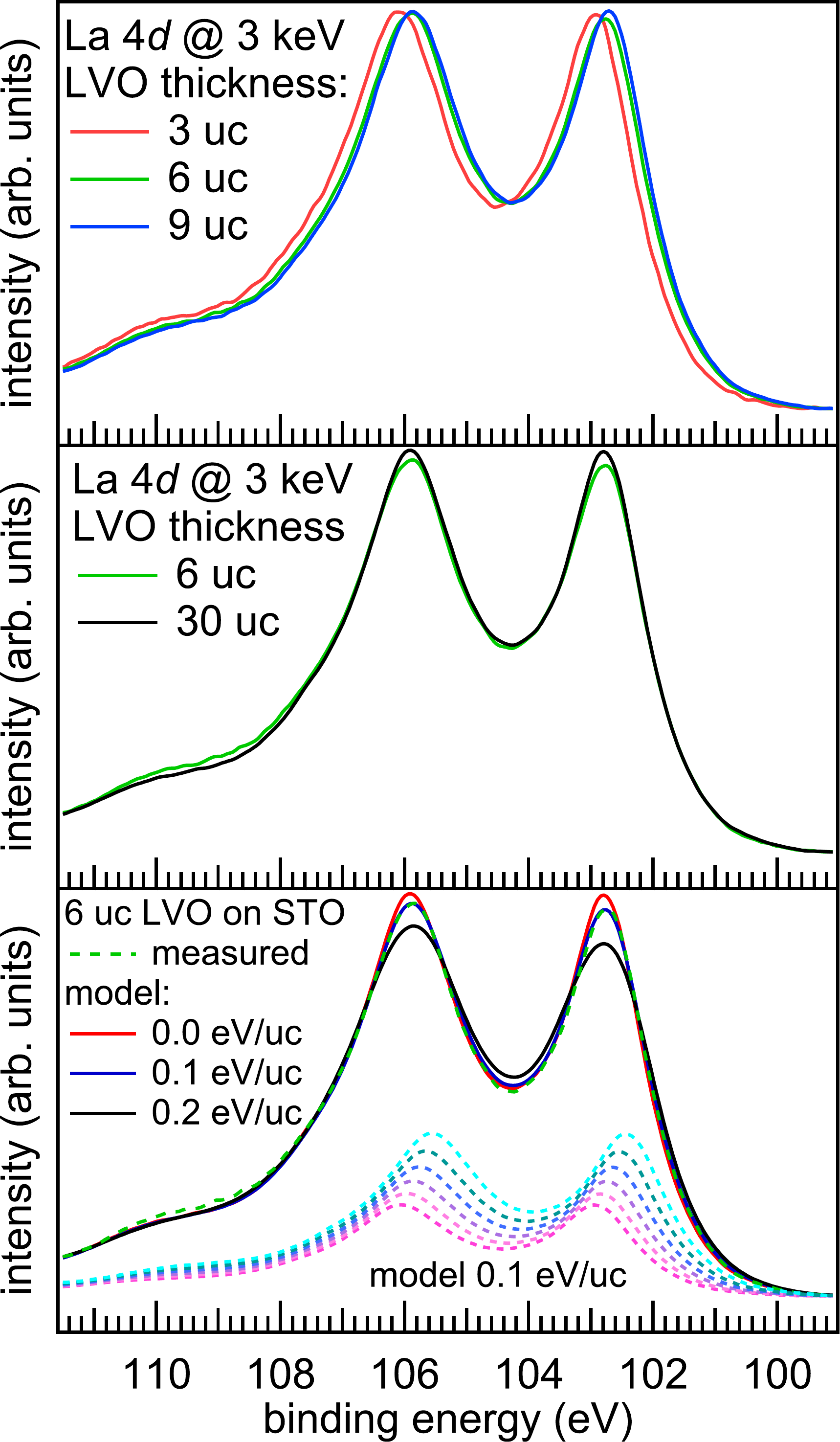}
\caption{
\label{monotonic}
(top) Angle-integrated La~4\textit{d} spectra for three 
different LVO/STO 
samples with LVO film thicknesses of 3, 6 and 9\,uc integrated over an emission 
angle $\theta$ of $(35\pm28)^\circ$. One can see an energy 
shift towards lower binding energies with increasing film thickness. (middle) 
Comparison of the La\,4\textit{d} spectrum of the film with a thickness of 
6\,uc with a reference spectrum of a 30\,uc thick film. The former is 
broader consistent with a potential gradient in the film. (bottom) 
Comparison of the spectrum of the 6\,uc LVO/STO sample with model spectra for three 
different potential gradients. The model spectrum based on a uniform gradient 
of 0.1\,eV/uc best matches the measured spectrum. Also shown is the 
decomposition of the model spectrum into its single components, corresponding 
to the six layers. Each single spectrum is exponentially damped according to 
the depth of its corresponding layer with its energy shift being determined by 
the gradient.}
\end{figure}

To gain insight into a possible electronic potential building up across the film, 
we measured three different LVO/STO samples with varying LVO thickness 
(3\,uc, 6\,uc and 9\,uc). For all film core-levels we notice a similar 
trend. With increasing film thickness, there is an increasing energy shift towards lower 
binding energies. Exemplarily, we show angle-integrated La~4\textit{d} 
core-level spectra in Fig.\,\ref{monotonic} (top). As can be clearly seen, the 
shift between the spectra of the 3\,uc and the 6\,uc films is bigger than 
that between the spectra of the 6\,uc and 9\,uc films. Note that these spectral 
shifts are a property of the LVO films as no shifts are seen in the substrate 
core-level spectra (not shown here). Such a monotonic shift of film core-level 
spectra with thickness can be a signature of a built-in potential, as has 
been reported, e.g., for LaCrO$_3$ and LaFeO$_3$ films grown on 
STO \cite{chambers_band_2011,comes_interface_2016}. In contrast, in a HAXPES 
study of LAO/STO, the film core level spectra did not 
shift with film thickness, indicating that there is no sizable potential 
gradient in this system \cite{berner_band_2013}.

To verify that there is indeed 
a built-in potential in the LVO film, we compare in Fig.~\ref{monotonic} 
(middle) the La~4\textit{d} spectrum of a 6\,uc LVO film with that of a 30\,uc thick 
reference sample. This reference should approximately have no potential 
gradient, which can be understood in an ideal electronic reconstruction picture: When assuming 
the highest occupied LVO band to cross the Fermi energy at a thickness of 
30\,uc instead of 6\,uc, the (average) potential gradient is a factor of five 
smaller. Indeed, the spectrum of the 6\,uc film is slightly broader than that of 
the thick one, which is consistent with a potential gradient in thinner films.

To obtain a rough estimate of the gradient, we simulate the measured 
La~4\textit{d} spectrum 
by the superposition of six La~4\textit{d} reference spectra measured from the 
30\,uc sample, one for each of the six film layers. Each reference spectrum is 
shifted in 
energy according to the potential gradient. The intensity $I$ from each layer $n$ is assumed to be damped according to 
\begin{align}
I(z=na)=I_{0} \exp(-na/{\lambda} \cos \theta),
\label{intensityeq}
\end{align}
where $I_{0}$ is the 
undamped signal, $\lambda$ the inelastic mean free path of the photoelectrons, $\theta$ the electron emission angle, $z$ the depth and $a$ the lattice constant of the film. The comparison between the 
measured 6\,uc spectrum and model spectra for three different gradients is 
depicted in Fig.~\ref{monotonic} (bottom). With such a 
coarse model, there is already a good match between the data and the simulated spectrum 
for a slope of 0.1\,eV/uc, whereas the other model spectra deviate 
more strongly from the measured spectrum, particularly near the peak maxima.    

\begin{figure*}[!htb]
\includegraphics[width = 0.99\linewidth]{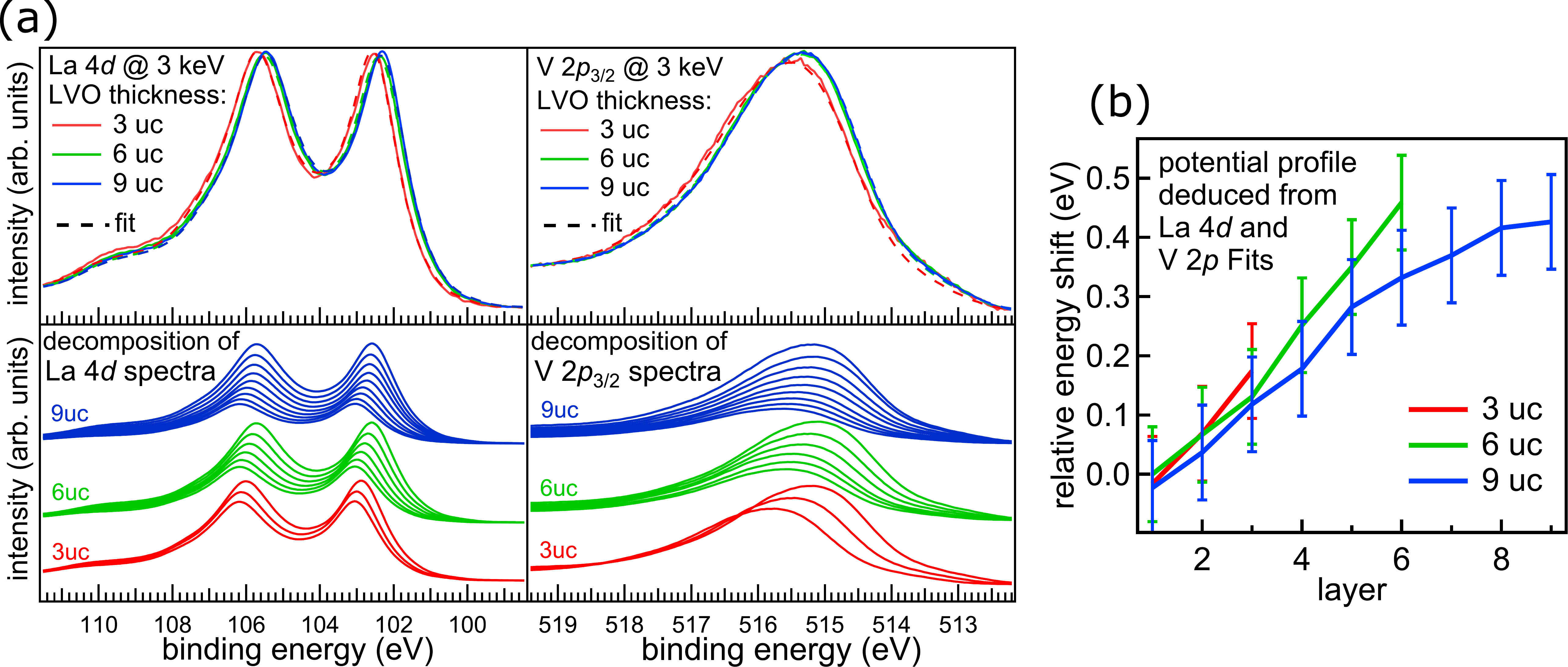}
\caption{
\label{Grad}
(a) top: La~4\textit{d} and V~2\textit{p}$_{3/2}$ spectra for 
three different LVO/STO samples with LVO film thicknesses of 3, 6 and 9\,uc 
along with the corresponding fit curves. For both core levels, one can see an 
energy shift towards lower binding energies with increasing film thickness. The 
fits match the measured spectra fairly well. (a) bottom: Decomposition of the 
La~4\textit{d} and V~2\textit{p} fit curves into their single spectra for 
all three thicknesses. Each single spectrum is exponentially damped according to 
the depth of its corresponding layer while the energy position is determined 
by the fitting routine. (b) The potential profiles for each film as deduced from the fitting of 
the La~4\textit{d} and V~2\textit{p} spectra in (a). The energy 
shift of the first layer of the 6\,uc sample at the interface (layer 1) 
is set to zero, with a positive shift indicating a shift to a lower binding 
energy.}
\end{figure*}

To refine our quantitative analysis of the potential 
gradient further, we \textit{fit} the measured La~4\textit{d} \textit{and} 
V~2\textit{p}$_{3/2}$ spectra at once for each of the three film thicknesses. 
The 
results are depicted in the top panels of Fig.~\ref{Grad} (a). As before, each measured core-level spectrum is built up from the spectra of the individual film layers. In contrast to before, the energy shifts across the film are no longer fixed but are fitted. The energy positions 
of the reference spectra are the only fitting parameters. The layer resolved spectra obtained from the best fits for 
all three samples are shown in the bottom panels of Fig.~\ref{Grad} (a). 
Altogether, one can see that the broadening and the 
thickness-dependent energy shifts of the measured film core-levels can be well 
reproduced by the fits with a significant potential gradient at work. From the 
energy shifts of the layer-resolved spectra for both La~4\textit{d} and 
V~2\textit{p}, we can deduce the potential profiles in the films, which are 
depicted in Fig.~\ref{Grad} (b). Here, the energy shifts of the 
individual spectra for all film layers are plotted relative to the 
spectrum of the first layer above the interface of the 6\,uc sample, with a 
positive energy shift indicating a shift to a lower binding energy. We obtain a 
potential gradient of around 0.1\,eV/uc for the 3\,uc and 
6\,uc films, while for the 9\,uc film the potential profile appears to level off toward the surface, 
resulting in a slightly lower average gradient. Furthermore, there is no 
significant potential offset between the three samples right at the interface.

In summary, we indeed find a sizable potential gradient, but it is 
significantly lower than the theoretical value of around 
0.3\,eV/uc \cite{assmann_oxide_2013}. One possible reason for this could lie in charged interface or 
surface defects or structural distortions, which partially 
screen out the gradient and have not been considered in the theoretical calculations \cite{assmann_oxide_2013}.

\subsection{Band bending in the SrTiO$_3$ substrate from core-level analysis}

\begin{figure*}[!htb]
\includegraphics[width = 0.85\textwidth]{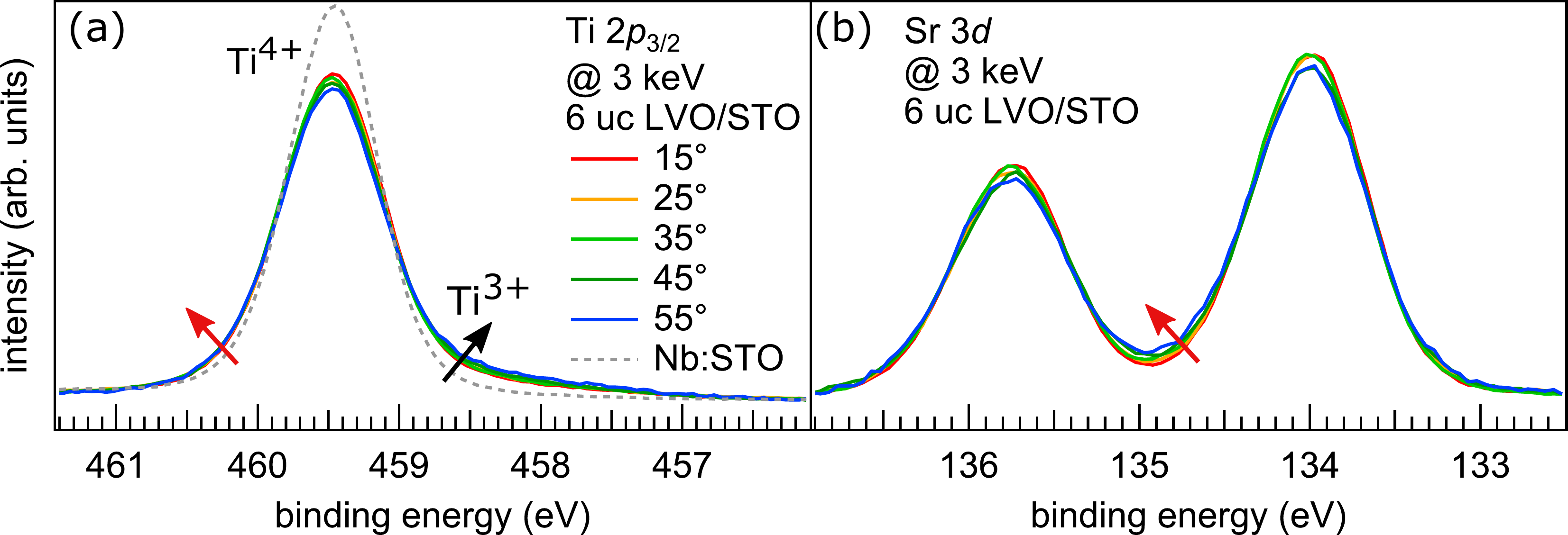}
\caption{
\label{coreTi}
Ti~2\textit{p}$_{3/2}$ (a) and Sr~3\textit{d} (b) spectra of a 6\,uc 
LVO/STO sample for different electron emission angles along with a 
Ti~2\textit{p}$_{3/2}$ reference spectrum of a Nb-doped STO substrate. Beside 
additional Ti$^{3+}$ spectral weight at lower binding energies (black arrow), 
both core level lines show an asymmetry towards higher binding 
energies as well (red arrows).}
\end{figure*}

\begin{figure}[!htb]
\includegraphics[width = 0.90\linewidth]{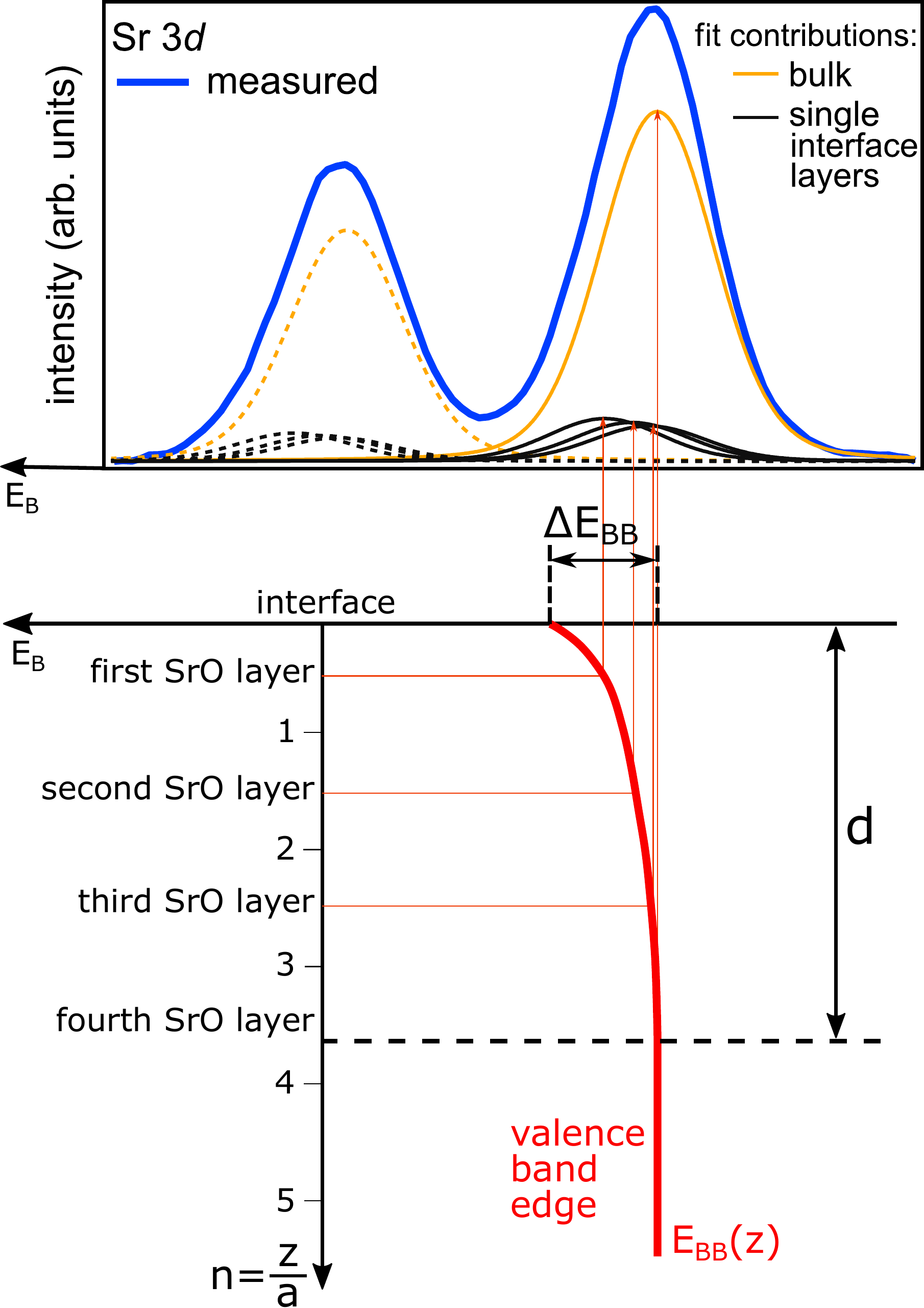}
\caption{
\label{BBexp}
Visualization of the interfacial band bending exemplified by a Sr~3\textit{d} 
spectrum. The photoemission signal of each SrO layer at depth \textit{z} is 
assumed to be a 
symmetric Voigt peak shifted in energy by $E_{bb}(z)$. The 
contributions from deeper layers without energy shift are treated as a single 
bulk peak. By 
assuming a quadratic potential within the bending zone, we can deduce the 
energetic depth of the band bending $\Delta E_{bb}$ and its spatial extension 
$d$.}
\end{figure}

\begin{figure*}[!htb]
\includegraphics[width = 0.95\textwidth]{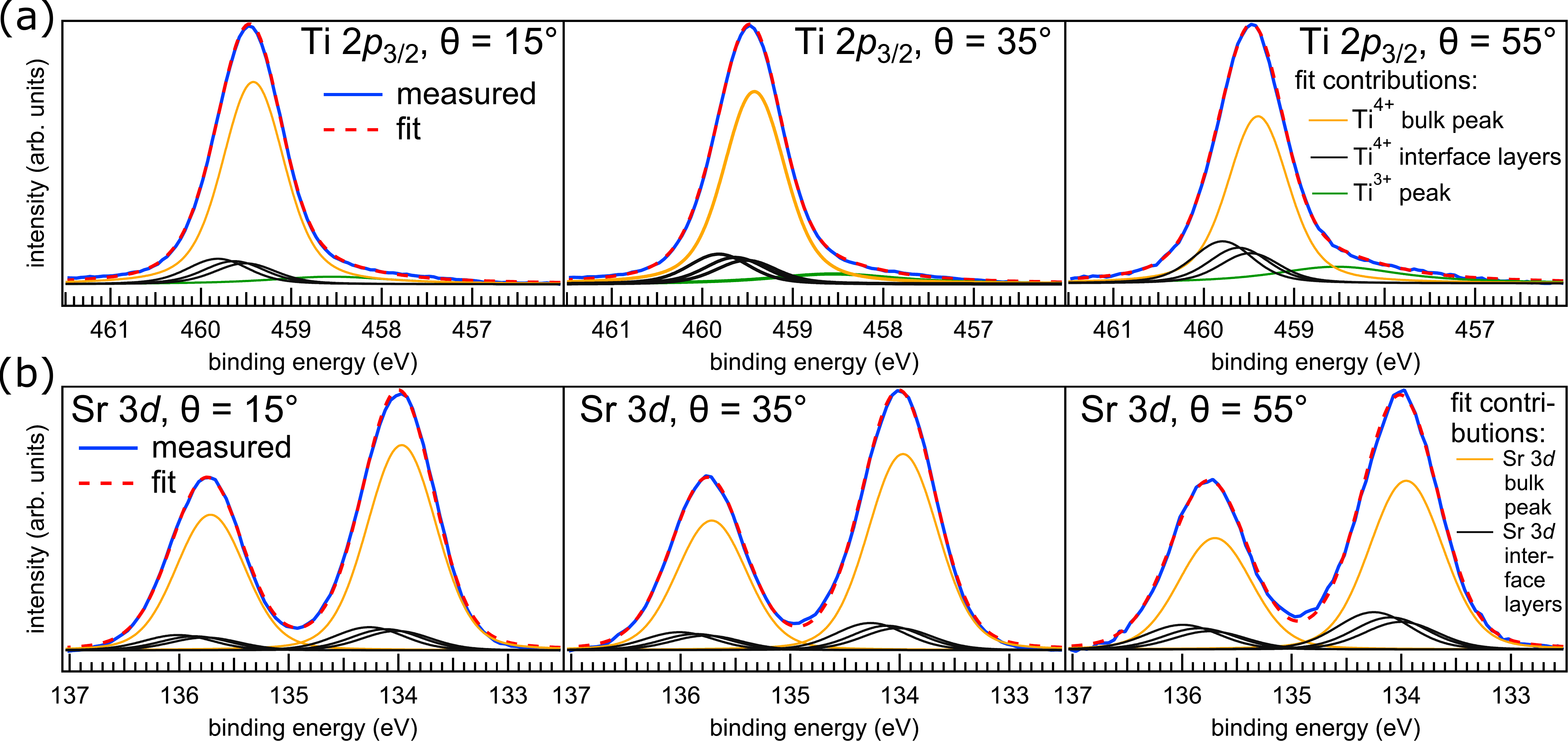}
\caption{
\label{globalTi}
(a) Exemplary fits for the Ti~2\textit{p}$_{3/2}$ photoemission spectra. The 
asymmetric line shape can be modeled by several, energetically shifted Voigt 
peaks, representing the individual layers in the bending zone, in addition to 
the Ti$^{4+}$ bulk peak. Beside the dominant Ti$^{4+}$ peak, there is an 
additional Ti$^{3+}$ contribution. (b) Exemplary fits for the Sr~3\textit{d} photoemission spectra. For each layer in 
the bending zone an energetically shifted spectrum is taken into account 
in addition to a single peak, representing the bulk contribution. }
\end{figure*}

Due to the enhanced probing depth of HAXPES, one can study not only the film 
but also the substrate core-levels. Figure~\ref{coreTi} depicts the 
Ti~2\textit{p}$_{3/2}$ and the Sr~3\textit{d} spectra of 
a 6\,uc LVO/STO sample recorded at 3\,keV photon energy for different electron 
emission 
angles $\theta$.  

Turning first to the Ti~2\textit{p}$_{3/2}$ spectra in 
Fig.~\ref{coreTi}, the corresponding Ti oxidation state is, by comparing with the 
reference spectrum of Nb-doped STO, readily identified 
as 4+. More interestingly, this comparison reveals additional spectral weight 
at low binding energies in the LVO/STO spectra, which is not present in the 
reference Nb:STO spectrum. This is known to originate from Ti$^{3+}$ ions which 
host an 
additional electron in the 3$d$ shell, leading to 
better screening of the core 
potential and hence a binding energy shift of the Ti~2\textit{p} peaks to lower 
values \cite{sing_profiling_2009}. As the spectral weight of the Ti$^{3+}$ shoulder
increases for higher emission angles $\theta$ (i.e. for higher interface 
sensitivity), these Ti$^{3+}$ ions reside mainly at the interface.

Apart from the Ti$^{3+}$ signal on the lower binding energy side, there is also 
a less pronounced broadening towards higher binding energies. 
Such a broadening can be even more 
clearly seen at the Sr~3\textit{d}$_{5/2}$ line, where one can also notice (see 
the red arrow) that the asymmetry becomes stronger towards higher electron 
emission angles, i.e., higher interface sensitivity. These 
asymmetries of core lines are ascribed to band bending, namely in 
the substrate towards the interface \cite{berner_band_2013,schutz_band_2015}. 
The binding 
energies of all substrate core levels follow the corresponding electronic 
potential profile. Each layer in the substrate bending zone contributes a 
photoemission spectrum that is shifted 
with respect to the bulk, with the 
total spectrum being a superposition of all these contributions, resulting in a 
broadened and asymmetric peak shape. Here, apparently a downward band bending 
towards the interface is observed as with increasing interface sensitivity the 
spectra originating from closer to the interface exhibit a shift 
to higher binding energies. 

To obtain quantitative estimates of the band bending, we introduce a model similar 
to Sch\"utz \textit{et al.} \cite{schutz_band_2015}. A visualization thereof 
is depicted in Fig.~\ref{BBexp}. We assume a potential profile of the 
form $E_{bb}(z)=\Delta E_{bb}(\frac{z}{d}-1)^2$  for $0 < z < d$ 
and $E_{bb}(z)=0$ elsewhere, with $E_{bb}(z)$ being the electrostatic potential at position $z$ below the interface, 
$\Delta E_{bb}$ the energetic depth of the band bending and $d$ the spatial extension of the bending zone. 
The photoemission signal from each layer of the substrate at depth \textit{z} 
is assumed 
to be a symmetric Voigt peak with its energy position shifted by $E_{bb}(z)$. 
The intensity from each layer is damped according to 
equation (\ref{intensityeq}). Note that the number of layers 
in the bending zone is not predetermined but a result of the fit. The 
contributions of all layers below the bending zone are summed up to constitute a single bulk peak.
To improve the statistical reliability of our analysis we fit simultaneously 
with this model the five Sr~3\textit{d} spectra and the five 
Ti~2\textit{p}$_{3/2}$ spectra collected at emission angles of
15$^{\circ}$, 25$^{\circ}$, 35$^{\circ}$, 45$^{\circ}$, and 55$^{\circ}$. By 
this 
global fit we obtain the values of $\Delta E_{bb}$ and $d$ with a total of 28 
fitting parameters \footnote{The number of parameters is narrowed down by 
physically reasonable constraints, e.g., each Sr~3\textit{d} peak has the same Gaussian and 
the same Lorentzian full width at half maximum for every angle.}. Figure~ 
\ref{globalTi}
exemplifies the fitting results for the 6\,uc LVO/STO sample. In general, the 
measured spectra are well reproduced by our model with negligible deviations 
supporting band bending being the main cause of the 
asymmetric peak shapes. Our quantitative analysis yields a value of 1.5\,nm for 
the spatial extension of the bending zone $d$ and about 0.4\,eV for the 
energetic depth of the band bending $\Delta E_{bb}$.

The fits in Fig.~\ref{globalTi} (a) also confirm that the Ti$^{3+}$ contribution 
rises for higher angles, reflecting the 
fact that Ti$^{3+}$ ions reside at the interface.

\subsection{Band offset between substrate and film from valence band analysis}

\begin{figure}[!htb]

\includegraphics[width = 0.99\linewidth]{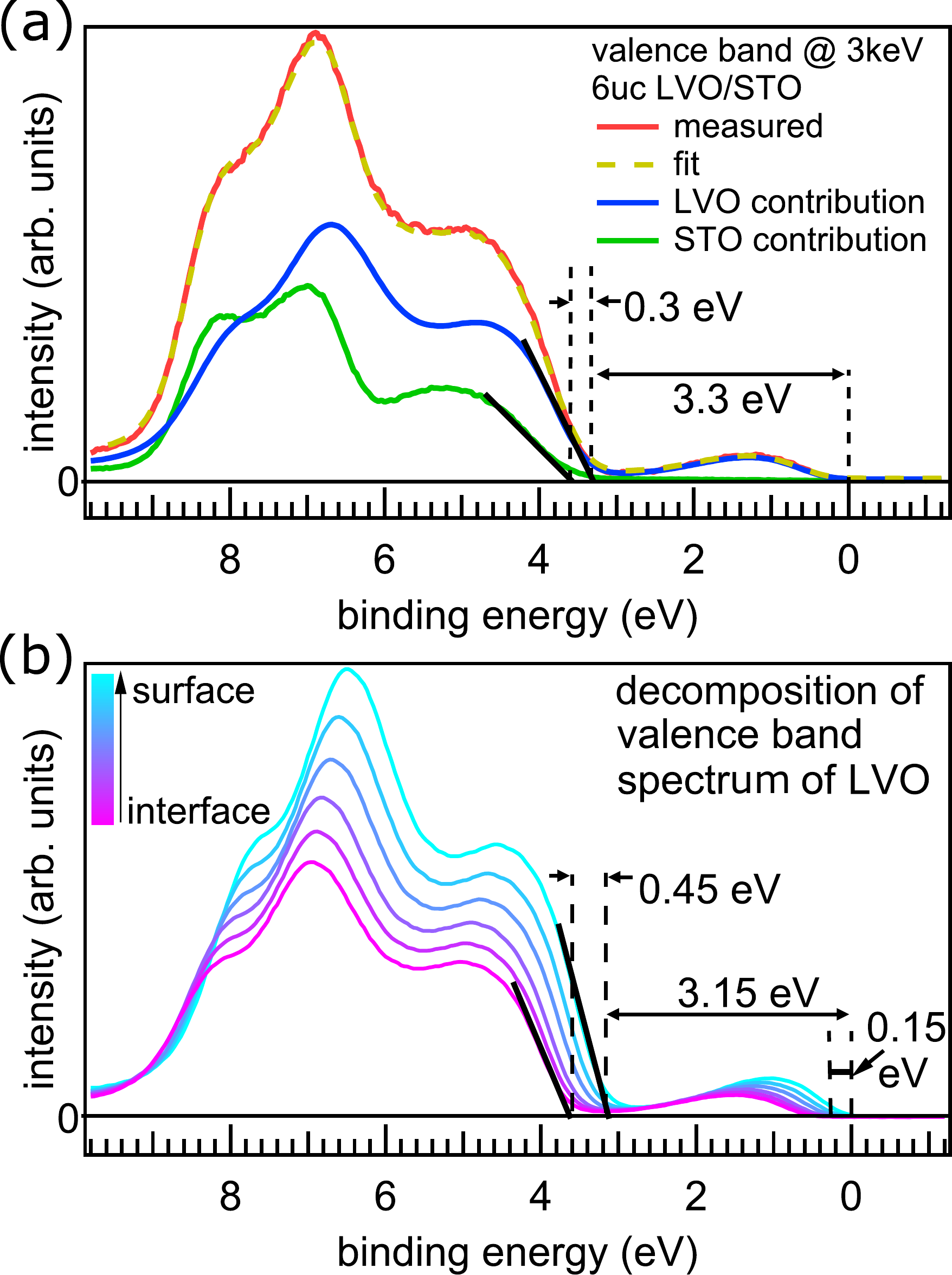}

\caption{
\label{VBfit}
(a) Valence band spectrum of a 6\,uc LVO/STO heterostructure and its 
decomposition into the individual LVO and
STO contributions. The valence band onsets are defined by linear 
extrapolations of the leading edges. (b) Layer-resolved decomposition 
of the LVO contribution from (a). The individual spectra are shifted in energy according 
to the potential profile.}
\end{figure}

A simple, accurate way to determine the band offset between substrate and film 
can be accomplished by analyzing the HAXPES valence band spectrum of 
a heterostructure \cite{berner_band_2013,giampietri_band_2017}, which is a weighted superposition of the valence 
band spectra of the film and substrate. The band offset corresponds to the energy difference between the leading 
edges of the valence band spectra of the film and substrate. To determine the band offset, we measured 
the reference valence band spectra from an STO substrate and a thick 
(30\,uc) LVO film, with which we fit the valence band 
spectrum of an LVO/STO heterostructure. The fitting parameters are the 
energy 
positions of the two spectra and their integral spectral weights. Figure 
\ref{VBfit} (a) depicts the valence band spectrum of 
a 6\,uc LVO/STO sample and its decomposition into the LVO and STO 
contributions. For the fitting, we included the potential profile in the LVO 
film shown in Fig.~\ref{Grad} (b). Hence, the LVO contribution is composed of 
the six layer-resolved LVO valence band spectra that are shifted in binding 
energy according to this profile, which are depicted in Fig.~\ref{VBfit} (b). 
The fitted spectrum is in very good agreement with the measured 
data. Note that there is a high contrast between the LVO and STO valence band 
spectra just below the Fermi energy where LVO shows spectral weight related to 
the lower Hubbard band whereas STO exhibits a gap. By comparing the binding 
energy difference between the valence band onsets of STO and LVO, which are 
determined by linear extrapolations of the leading edges to where zero 
intensity occurs, we obtain a valence band offset of about 0.3\,eV. However, 
since the LVO contribution is the sum of the single layers, the value for 
valence band onset, on which this analysis is based, 
is kind of averaged over the film. To derive a more precise value for the 
band offset at the interface, we compare the onsets of the valence band spectra of the bottom LVO layer [Fig.~\ref{VBfit}~(b)] and of the STO (bulk) contribution which yields 0. The onsets of the following layers 
track the potential gradient and hence shift to lower binding energies. The energy values depicted here are also derived from linear extrapolations of the leading edges. In particular, the valence band analysis allows us to complete our band scheme: The onset of the LHB of the top LVO layer is 0.15\,eV below the Fermi energy. With an energy difference of 0.45\,eV between the spectra of the bottom and top LVO layers, the onset of the LHB of the bottom layer is hence 0.6\,eV below $E_F$.

We note that due to the large inelastic mean free path compared to the bending 
zone extension in STO the contribution of bulk STO to 
the valence band spectrum by far outweighs that of the bending zone. 
In addition, in the bending region the spectral onsets of the 
layer-resolved signals shift downwards. Consequently, the band bending in STO can be ignored in determining the valence band edge of STO. 

\begin{figure}[!htb]
\includegraphics[width = 0.99\linewidth]{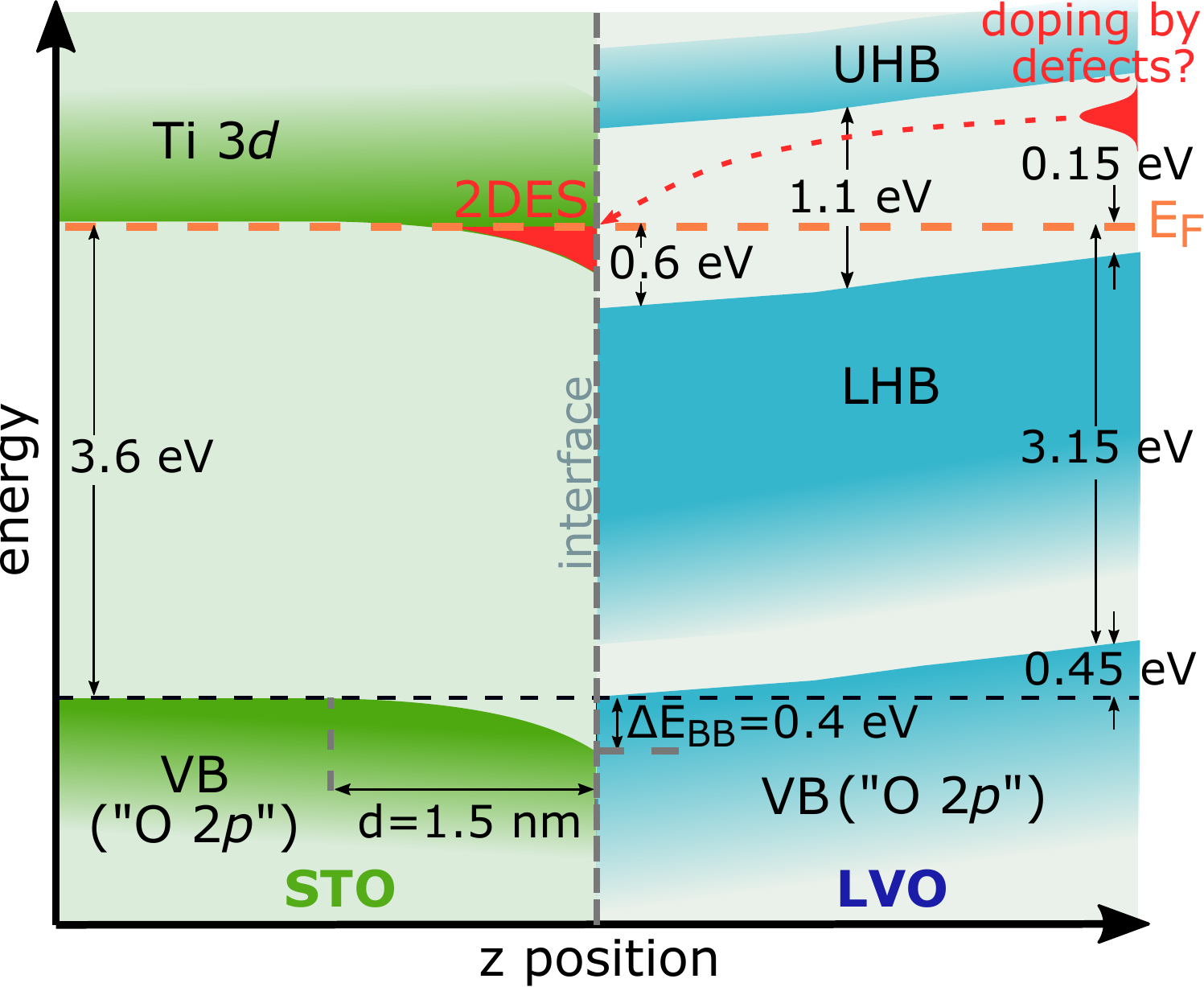}

\caption{
\label{VB}
Band diagram of the 6\,uc LVO/STO heterostructure as deduced from core level 
and valence band photoemission, with the value of the optical gap - as defined by the energy difference between the onsets of lower (LHB) and upper Hubbard bands (UHB) - taken from 
literature. In the substrate, the bands are bending downwards towards the 
interface while there is an increasing potential slope in the film. The 
Ti~3\textit{d} bands, crossing the Fermi level, provide a pocket 
for electrons to form a two-dimensional electron system (2DES) at the 
interface. The LHB of the LVO film does not quite reach the Fermi level at the film surface. The bulk valence band offset is zero, since the valence band states of LVO and bulk STO, denoted by VB ("O~2\textit{p}")\footnote{We note that for the energy regime of HAXPES, this part
of the valence band contains contributions not only from oxygen but also to a significant amount from the A and B cations in film and substrate \cite{thiess_valence_2010}. }, are aligned.}
\end{figure}

\subsection{Complete band diagram and summarizing discussion}

Having determined the potential profile in LVO, the band bending of STO near 
the interface and the interfacial band offset, and referring to the 
reported optical gap of LVO of 1.1\,eV \cite{wang_device_2015, arima_variation_1993}, we are able 
to derive the full electronic band diagram of the 6\,uc LVO/STO heterostructure, 
as drawn in Fig.~\ref{VB}. 

In a region extending from the interface into the substrate by about 1.5\,nm, 
the Ti~3$d$ conduction bands are bent downwards below the Fermi energy when 
approaching the interface. These states are filled by electrons, giving rise to 
a 2DES. The VB maxima of bulk STO and 
the first LVO layer at the interface are aligned, yielding a 
zero interfacial band offset. In the LVO film, there is a built-in potential due 
to the polar 
discontinuity which, however, is not large enough to shift the lowest lying 
excitations of the LHB up to the Fermi energy. The upper Hubbard band, in turn, 
lies way above the Fermi energy. Thus, no doping of V~3$d$-derived 
states takes place. 

Our transport and photoemission experiments conclusively show that the 
conductivity in LVO/STO heterostructures above a critical thickness of 5\,uc is 
generated by the intrinsic mechanism of electronic reconstruction, with 
Ti~3\textit{d} interface states hosting mobile electrons. Self-doping 
within the LVO film can be excluded from the experimentally derived band 
diagram as well as vanadium core level spectra and ResPES measurements of 
the valence band, showing no indication for a change 
of the vanadium valence state throughout the film and V~3$d$ spectral weight 
at $E_F$, respectively. Hence, a band-filling induced 
Mott-insulator-to-metal transition in LVO can be ruled out.

We find that the built-in 
potential is much lower than what is expected in the ideal picture of 
electronic reconstruction, based on density-functional calculations. 
Furthermore, the built-in potential does not cause a Fermi level crossing of occupied 
bands, in particular the LHB, in the film with increasing thickness. 
These observations can be reconciled in a refined picture of electronic 
reconstruction, as has been suggested for LAO/STO \cite{berner_direct_2013}. In this scenario, 
the interfacial charges do not stem from bands crossing the Fermi level 
but from defect states, in particular oxygen vacancies at the very film 
surface \cite{bristowe_surface_2011,li_formation_2011,pavlenko_oxygen_2012, 
yu_polarity-induced_2014-1,zhong_polarity-induced_2010}. Although it costs energy 
to build these defects, at the critical thickness, the energy gain by 
transferring the released charges to the interface and thus 
partially compensating the polarization field predominates. Hence, for 
the system studied here, we suggest surface oxygen defects as charge 
reservoir for the interface 2DES.

Knowledge of the complete band diagram as derived in this study can be very 
useful for developing novel concepts for photovoltaic devices. Particularly, the advantages of the 
intrinsic electric field in LVO/STO can be combined with the benefits of the 
optical gap of LVO, which is in the ideal range for absorption of visible 
light. The measured potential gradient in the LVO film could help to separate 
photogenerated carriers without the need of creating a $pn$-junction by means of 
chemical doping. In addition, the induced metallic interface reported here can 
be used as a contact for extracting these carriers. As we find a sizable 
gradient only for the first few layers and a gradient saturation for thicker films, one could think of LVO/STO 
superlattices with optimized LVO layer thicknesses just at or slightly above 
the critical value to maximize the potential build-up and facilitate the charge carrier separation. With 
suitable contacts on the metallic STO layers for optimized multiple carrier 
extraction, the photovoltaic property of the whole system can be greatly 
enhanced. Furthermore, it has been proposed to combine the LVO film with another material with a band gap in the solar range, 
e.g. LVO and LaFeO$_3$ \cite{assmann_oxide_2013}. 
With such band-gap graded designs the solar absorption can be increased even 
more by improving the conversion efficiency in the energy regions of the 
respective band gaps \cite{sassi_theoretical_1983, konagai_gradedbandgap_1975}. 
The materials combination of LVO and STO (possibly combined with additional 
solar absorbing materials) may hence allow the superior 
properties of Mott materials, in terms of quantum efficiency and energy 
conversion, to be exploited for photovoltaic applications.

\section{Summary}

Based on electrical transport and photoemission experiments we elucidate the 
mechanism leading to conductivity in the LaVO$_3$/SrTiO$_3$ heterostructure and 
the nature of the charge carriers. We find a critical film thickness of 5\,uc 
for the onset of metallicity. The comparably low charge carrier concentration 
of less than 10$^{14}$\,cm$^{-2}$ and the fact that its temperature dependence 
is almost identical to that of post-annealed LaAlO$_3$/SrTiO$_3$ leads us to 
the 
conclusion that conductivity is not induced by oxygen vacancies in the substrate 
but intrinsic in nature, viz., due to electronic reconstruction as a 
consequence of the polar discontinuity at the interface. From the 
photoemission spectra we infer a potential gradient in the LVO film of around 0.1\,eV/uc for 
films up to 6\,uc and a saturating behavior for larger thicknesses. In 
the substrate we find a downwards band bending with a spatial 
extension of about 1.5\,nm and an energetic depth of about 0.4\,eV. The 
conducting electrons reside in interfacial Ti~3$d$ 
states whereas no indications are found for occupied V~3\textit{d} states. We 
suggest surface oxygen vacancies as charge reservoir for the conducting 
interface.

The complete band diagram as derived from hard X-ray photoemission 
spectroscopy in combination with resistivity and Hall effect data in this 
study promote a better understanding of the LaVO$_3$/SrTiO$_3$ heterostructure. 
In particular, our results can help to harness the peculiar properties of Mott 
materials for tailored photovoltaic applications in future devices.

{\acknowledgments}
The authors thank G. Sangiovanni for helpful discussions. The authors are grateful for funding support from the Deutsche 
Forschungsgemeinschaft (DFG, German Research Foundation) under Research grant SI 851/2-1 (Project ID 431448015) and through the 
W\"urzburg-Dresden Cluster of Excellence on Complexity and Topology in Quantum 
Matter ct.qmat (EXC 2147, Project ID 390858490) as well as through the 
Collaborative Research Center SFB 1170 ToCoTronics (Project ID 258499086).
The authors acknowledge Diamond Light Source for time on beamline I09 under proposals SI17499 and SI23737.
The authors would also like to thank D. McCue for technical support at 
beamline I09.

\end{document}